\documentclass[12pt]{article}
\usepackage{amsmath}
\usepackage{amssymb}
\usepackage{epsfig}
\usepackage{euscript}
\usepackage{fancybox}
\usepackage{color}
\def\mathswitchr#1{\relax\ifmmode{\mathrm{#1}}\else$\mathrm{#1}$\fi}

\newcommand {\pslash}{\hbox{$\not\hbox{\kern-2.3pt $p$}$}}

\usepackage{cite}

\usepackage{epic}
\def\alf1{ {\alpha\over\pi} }

\begin{document}
%\input{feynman} 
%=======================================================================
\begin{titlepage}
\begin{flushright}
%{\bf CERN-PH-TH/2011-077}\\
{\bf BU-HEPP-11-05}\\
{\bf Dec., 2011}\\
\end{flushright}
\vspace{0.05cm}
 
\begin{center}
{\Large Estimates of Radiation by Superluminal Neutrinos}
\end{center}

\vspace{2mm}
\begin{center}
%%  {\bf   S. Jadach$^{a,b}$ and B.F.L. Ward$^{c,d}$}
{\bf   B.F.L. Ward}\\
\vspace{2mm}
%{\em $^a$CERN, Theory Division, CH-1211 Geneva 23, Switzerland,}\\
%{\em $^b$Institute of Nuclear Physics,
%        ul. Kawiory 26a, Krak\'ow, Poland,}
%{\em $^c$Werner-Heisenberg-Institut, Max-Planck-Institut fuer Physik,
%Muenchen, Germany,}\\
%{\em $^a$Werner-Heisenberg-Institut, Max-Planck-Institut fuer Physik,
%Muenchen, Germany,}\\
{\em Department of Physics, Baylor University, Waco, TX, USA}\\
%{\em $^b$ Department of Theoretical Physics, IACS, Kolkata, India}\\
%{\em $^c$Department of Physics, The Citadel, Charleston, SC, USA}\\
%{\em TH Unit, CERN, Geneva, Switzerland}\\
%{\em $^c$SLAC, Stanford University, Stanford, California 94309, USA,}\\
%{\em $^b$Department of Physics and Astronomy,\\
%  The University of Tennessee, Knoxville, Tennessee 37996-1200, USA}\\
%{\em $^c$SLAC, Stanford University, Stanford, California 94309, USA,}\\
\end{center}

\vspace{5mm}
\begin{center}
{\bf   Abstract}
\end{center}
We show that the more
energetic superluminal neutrinos with quadratically dispersed
superluminalities $\delta=\beta^2-1$, for $\beta=v/c$ where $v$ is the 
neutrino velocity, also lose significant energy to radiation to the 
$\nu+e^-+e^+$ final state in traveling from CERN 
to Gran Sasso as has been shown to occur for those with constant 
superluminality by Cohen and Glashow 
if indeed $\delta \simeq 5\times 10^{-5}$. 
In addition, we clarify the dependence 
of such radiative processes on the size of the superluminality. 
\vskip 3mm

\vskip 16mm
\vspace{10mm}
\renewcommand{\baselinestretch}{0.1}
\footnoterule
\noindent
%{\footnotesize
%\begin{itemize}
%\item[${\dagger}$]
%Work partly supported by US DOE grant DE-FG02-09ER41600. 
% the Polish Government
%grants KBN 2P30225206 and 2P03B17210, the Maria Sk\l{}odowska-Curie
%Joint Fund II PAA/DOE-97-316, and
%by NATO Grant PST.CLG.980342.
%, and by
%Polish Government grant 5P03B09320.
%\end{itemize}
%}
%\vspace{0.5cm}
%\begin{flushleft}
%{\bf UTHEP-00-0101}\\
%{\bf Jan, 2000}\\
%\end{flushleft}

\end{titlepage}
 
\baselineskip=11pt 
%=======================================================================
\def\Kmax{K_{\rm max}}\def\ieps{{i\epsilon}}\def\rQCD{{\rm QCD}}
\renewcommand{\theequation}{\arabic{equation}}
\font\fortssbx=cmssbx10 scaled \magstep2
\renewcommand\thepage{}
%\vfill\eject
\parskip.1truein\parindent=20pt\pagenumbering{arabic}\par

%\section{\bf Introduction}\label{intro}\par
In trying to understand the results of the OPERA Collaboration~\cite{opera} 
for apparent faster than light velocity for neutrinos in transit from CERN 
to Gran Sasso, Cohen and Glashow~\cite{cg} have made an estimate of the 
Cherenkov effect to the $\nu+e^++e^-$ final state using the standard 
methods adapted to Lorentz
invariance-violating interacting quantum field theory. This effect is also 
studied in Ref.~\cite{matt} in a general framework for such violations of 
Lorentz invariance with particular attention to the case where the 
superluminality is quadratically dispersed in contrast to the case considered 
by Cohen and Glashow with constant superluminality, where we define the 
superluminality by 
$\delta=\beta^2-1$ when $v$ is the neutrino velocity. Here, we extend the 
analysis of Cohen and Glashow to the quadratically dispersed case of 
Ref.~\cite{matt} and we compare the analyses of Cohen and Glashow and 
of Ref.~\cite{matt} 
with an eye as well toward the very interesting observations of the 
OPERA Collaboration.\par
We need to first state very clearly at the outset here that we do not claim
to rule out the possibilities discussed in Refs.~\cite{dsr1a,dsr1b,carcor} in which appropriate deformation
of the Lorentz symmetry renders these Cherenkov processes kinematically 
forbidden just as one has for the usual non-superluminal case in the usual special theory of relativity. Thus, non-observation of the effects discussed herein does not rule-out such changes in the standard orthodoxy.\par
Specifically, 
%characterizing the excess speed, presumed not to depend very much on energy (we return to this below), of the neutrinos by the parameter
%$\delta=\beta^2-1$, where as usual $\beta=v/c$, 
Cohen and Glashow arrive at the results
\begin{equation}
\begin{split}
\Gamma&=k'\frac{G_F^2}{192\pi^3}E^5\delta^3\\
\frac{dE}{dx}&=-k\frac{G_F^2}{192\pi^3}E^6\delta^3
\end{split}
\label{cg1}
\end{equation}  
for the Cherenkov radiation rate and energy loss per unit distance for the process $\nu\rightarrow \nu+e^++e^-$ for $\nu$'s of energy $E$ where the parameters $k',\;k$ 
are $1/14$ and
$25/448$, respectively~\cite{dnan}. In this way, 
assuming that $\delta$ is independent of $E$,
they find~\cite{cg} that the $\nu$'s should lose considerable energy between
CERN and Gran Sasso.\par 
On the other hand, Mattingly {\em et al.} in Ref.~\cite{matt} 
have discussed the same
transition for superluminal $\nu$'s via analogy with the 
so-called $\nu$-splitting process $\nu_A\rightarrow \nu_A\nu_B\bar\nu_B$ for which they find the result
\begin{equation}
\Gamma_{\nu\nu\bar\nu}\sim \frac{3G_F^2}{64\pi^3}\frac{\eta_\nu^4E^{13}}{M_{Pl}^8}
\label{mattg1}
\end{equation}
where the factor of $3$ multiplying $G_F^2$ sums over three neutrino species
so, ignoring the masses of the neutrinos and the electron, Mattingly {\em et al.}
assert (see Sect. 6 of Ref.~\cite{matt}) that the rate for the $\nu+e^+e^-$ decay should be nearly the same as that for the $\nu$-splitting decay in (\ref{mattg1})\footnote{The sum over the three neutrino species mildly over compensates for the almost equal coupling of the Z$^0$ boson to left and right handed electrons so that the rates would become almost equal if we multiply latter one by 2/3 -- we ignore such factors here.}. The Lorentz-violating
parameter $\eta_\nu$ is such that the dispersion relation for the $\nu$
is $$E_p^2=m_\nu^2+p^2+\eta_\nu p^4/M_{Pl}^2$$
where $M_{Pl}$ is the Planck mass, $m_\nu$ is the neutrino rest 
mass and $p$ is the $\nu$ momentum. The parameter $\eta_\nu$ 
is to be determined
from experiment.
Thus, using $v=\frac{dE_p}{dp}$ and ignoring the neutrino mass 
we see that $$\delta\cong 3\eta_\nu E^2/M_{Pl}^2$$
so that Mattingly {\em et al.} argue that the rate for the $\nu\rightarrow \nu+e^++e^-$process should be nearly given by
\begin{equation}
\begin{split}
\Gamma_{\nu e^+e^-}&\simeq \Gamma_{\nu\nu\bar\nu}\\
      &\cong \frac{3G_F^2}{64\pi^3}\frac{\eta_\nu^4E^{13}}{M_{Pl}^8}\\                  & \cong \frac{3G_F^2}{64(81)\pi^3}E^5\delta^4.
\end{split}
\label{mattg2}
\end{equation}
While the numerical factors and the explicit dependence on $E$
are consistent between the two estimates of $\Gamma$
for the $\nu\rightarrow \nu+e^++e^-$ process, the two dependencies on $\delta$
differ by one factor $\delta$: this factor would obviate the arguments
of Cohen and Glashow, given that OPERA have found that $\delta\cong 5\times 10^{-5}$. Which dependence on $\delta$ is actually correct?
The literature is beginning to feature use of both dependencies, as we see
in Ref.~\cite{carcor} where the corresponding dependence of Mattingly {\em et 
al.} is used and in Ref.~\cite{matt2} where both dependencies are featured
\footnote{By going to the 'preferred frame' of the incident $\nu$, we see 
from the textbook results in Ref.~\cite{bj-d} that, when we neglect all final 
state particle masses, the total rate to the $\nu e^-e^+$
final state is in fact independent of the handedness of the outgoing electron 
when $v_e$, the SM vector coupling of $e^-$ to the $Z^0$, is set to zero. 
Thus the arguments
in Ref.~\cite{matt2} as to why their splitting process rate has a different 
dependence on $\delta$ are not valid.}. Secondly, what effect does quadratic 
dispersion for $\delta$ have on the results of Cohen and Glashow? 
We address these two questions here in inverse order.\par
We see that, unlike the case discussed by Cohen and Glashow, we have  
to consider 
pronounced dependence of $\delta$ on momentum: for a neutrino of initial 
energy $E_0\sim 20$GeV, $\delta(E)\cong\delta_0(E/E_0)^2$ in an obvious 
notation. Moreover, the factor of 3 between the deviation between 
the squared group velocity and 1 here
and the squared incoming $\nu$ four-momentum vector in units of its squared
laboratory
energy $E$ means that we should identify  the quantity $\delta$
in the Cohen-Glashow formulas in (\ref{cg1}) with our $\delta/3$.
Thus, the dependence of $dE/dx$ on $E$ here is given by
\begin{equation}
\frac{dE}{dx}=-k\frac{G_F^2}{192\pi^3}\frac{E^{12}\delta_0^3}{E_0^6}.
\label{eqnw1}
\end{equation}
Integrating this relation gives us
\begin{equation}
\begin{split}
\frac{1}{E^{11}}-\frac{1}{E_0^{11}}&=11k\frac{G_F^2}{192\pi^3}\frac{\delta_0^3L\
}{E_0^6}\\
&=\frac{11}{5}\frac{1}{E_T^5E_0^6}
\end{split}
\label{eqnew2}
\end{equation}
where we introduced the distance $L\cong730$km from CERN to Gran Sasso
and the Cohen-Glashow quantity 
${1/E_T^5}\equiv \frac{5kG_F^2\delta_0^3L}{192\pi^3}$ 
so that for $\delta_0=5\times 10^{-5}$ $E_T\cong 12.8GeV$. 
This should be
compared with Eq.(21) in Ref.~\cite{matt2}, wherein only
one power of the analog
of $\eta_\nu$ appears versus three powers here.
If we define the average value of $\delta$ via\footnote{Note that the strict 
definition of the average velocity, $<v>=\int dt v(t)/T\equiv L/\int dx/v(t)$, 
reduces to
this result in (\ref{eqv1}) in the limit in which we work wherein 
all velocities are close to 1.}
\begin{equation}
\begin{split}
<\delta>&=\frac{1}{L}\int_0^Ldx \delta(x)\\
        &= \frac{1}{L}\int_0^Ldx (E^2(x)/E_0^2)\delta_0,
\end{split}
\label{eqv1}
\end{equation}
we arrive at the relation
\begin{equation}
<\delta>=\frac{5\delta_0}{9}\frac{E_T^5}{E_0^5}\left(\left(1+\frac{11}{5}\frac{\
E_0^5}{E_T^5}\right)^{\frac{9}{11}}-1\right).
\label{eqv2}
\end{equation}
This means that, if we identity $3<\delta>$ with the OPERA result 
$5\times 10^{-5}$, we can solve this latter relation (\ref{eqv2})
to get $3\delta_0=5.3\times 10^{-5}$ when the initial energy is $E_0=20$GeV.
Thus, even though $\delta$ varies quadratically with momentum, it does
not deviate very much from its initial value and a $20$ GeV $\nu$
now loses about 5.8\% of its energy in travelling from CERN to Gran Sasso
%\footnote{This is consistent 
%with the more qualitative treatment in Ref.~\cite{rao} which finds
%that a $17.5$GeV OPERA neutrino would lose about 4.2\% of its energy 
%in making the same trip with a quadratically dispersed value of $\delta$ --
%our calculation would yield a 3.2\% loss for this case.},
compared to 37.4\% for the case discussed by Cohen and Glashow.
The two groups of average velocity data in Ref.~\cite{opera} have average 
energies 13.8 and 40.7 GeV with superluminalities of $\beta-1$ given 
by $(2.25\pm .81)\times 10^{-5}$ and $(2.79^{+.84}_{-.83})\times 10^{-5}$
respectively. Solving for the respective values of
$3\delta_0$ we get $5.1\times 10^{-5}$ and $1.2\times 10^{-4}$
with energy losses of 1.1\% and 41.8\% respectively, where we use again
$3<\delta>\cong 5\times 10^{-5}$ in view of the errors on the data
without loss of content -- the respective energy losses in the
Cohen-Glashow analysis are 16.6\% and 68.6\%. The factor of order
2 variation of $\delta_0$ with initial creation energy is not excluded
by the data. In any case, when the $\nu$ energy is in the 10 MeV regime,
our values of $\delta$ are well below the limits from SN1987a~\cite{sn1987a},
$4\times 10^{-9}$,
due to the quadratic dependence on energy in principle. We say 'in principle' 
because there is, presuming the OPERA result is true, an unknown source of
this energy dependence and possibly
of the variation of $\delta_0$ with initial energy
and we cannot make definitive statements until this source is known.
We need more experimental input.\par
Concerning the detailed relationship between our calculation and that in Ref.~\cite{rao}, we note that the authors in Ref.~\cite{rao} make a numerical estimate for the 
$\nu\rightarrow \nu +e^++e^-$ process with the quadratic dispersion for the 
$\nu$ superluminality and with the usual special relativity dispersion for the $e^-,\; e^+$
with the result that the coefficient
of $G_F^2E^5\delta^3(1-4\sin^2\theta_w+8\sin^4\theta_w)$ in the Cohen-Glashow process is 1/45197.8 vs the exact result~\cite{cg,dnan} 1/41672.4, neglecting the lepton rest masses. This is acceptable in view of the current errors on the data. The authors in Ref.~\cite{rao} then approximate the energy loss using a constant value of $\delta$ so that they give in their eq.(21) a mean decay time that scales as the 5th power of the $\nu$ energy when in fact the dependence of $\delta$ on energy shows that the decay time has to scale with the 11th power of this energy, as we have shown here. This results in significant deviation from the correct decay energy loss profile as we have shown above for the higher energies in the OPERA spectrum. However, for the 17.5 GeV energy $\nu$'s, 
we find that they lose
3.2\% of their energy in travelling from CERN to Gran Sasso whereas the authors in Ref.~\cite{rao} estimate this as a 4.2\% loss if you take their result literally as written below their eq.(20). If you interpret their 4.2\% as the mean number that decay via the $\nu\rightarrow \nu+e^-+e^+$ process so that we do not actually lose $\nu$'s but we lose their energy, then to get an estimate of the corresponding energy loss we have to multiply by the mean energy loss per decay, which is ~\cite{cg,dnan} 78\%. This would then agree 
with our 3.2\% result -- this is expected because as the energy does not change very much the assumption of a constant $\delta$ is not a bad one. We note as well that the authors in Ref.~\cite{rao} do not address the issue of the
difference between the power of $\delta$ in the decay rate formulas in (\ref{cg1}) on the one hand and in (\ref{mattg1}) and (\ref{mattg2}) on the other. We now turn to this issue.\par
To answer the question on the proper power law for the dependence
of the radiation rate on $\delta$, we revisit the analysis in
Ref.~\cite{matt} and observe that, in their Eq.(4.10), there is an 
energy conserving delta function, $\delta(E_p-E'_p-E_q-E'_q)$, where for definiteness we reproduce this equation here as 
\begin{equation}
\Gamma_{AB}=\frac{g^4}{16(2\pi)^5M_Z^4cos^4\theta_w}\int\delta(E_p-E'_p-E_q-E'_q)d^3p'd^3qF^2
\end{equation}
for the width for the $\nu_A(p)\rightarrow \nu_A(p')\nu_B(q')\bar\nu_B(q)$ splitting process in a standard notation, with $E'_a=E_{a'},\; a=p,\; q$. Here, $q'(q)$ is the 4-momentum
of the $\nu_B(\bar\nu_B)$, respectively, and the $SU_{2L}\times U_1$~\cite{gsw} couplings
and parameters are standard by now -- the rest mass of the $Z^0$ is $M_Z$, the $SU_{2L}$ coupling constant is $g$, the Weinberg angle is $\theta_w$. 
The function $F$ contains the dependence of the differential decay rate on the
various angular phase space variables~\cite{matt}.
Upon arguing correctly that the 
excess energy from Lorentz-violation goes to transverse motion, Mattingly {\it et al.} then
estimate the size of this transverse phase space and argue that the only effect of the delta function constraint on energy is to remove one factor of the 
energy from their result for the integral over the two phase spaces for $p'$ and $q$. However, there is a Jacobian required to implement this removal, by the standard methods, given by the absolute value of the derivative of the argument of the delta function with respect to
$p'_{\parallel}$ for example\footnote{Here, $\parallel$ denotes the direction along that of the 3-momentum of the initial state $\nu$.}
due to the result
\begin{equation}
\delta(f(x))=\sum_{x_a\in\{\text{zeros of {\it f}}\}}\frac{\delta(x-x_a)}{|\frac{df(x_a)}{dx}|} 
\end{equation}
for any sufficiently well-behaved function $f(x)$
: This provides the missing factor of $\delta$ that should divide the expressions on the RHS of (\ref{mattg2}) and of (\ref{mattg1}) to bring them into agreement with Cohen and Glashow. To see this note that the absolute value of the derivative of the argument of the delta function 
with respect to $p'_\parallel$ can be represented as follows:
\begin{equation}
\begin{split}
\left|\frac{d(E_p-E'_p-E_q-E'_q)}{dp'_\parallel}\right|&=\left|\frac{d(E'_p+E'_q)}{dp'_\parallel}\right|\\
&=\left|(p'_\parallel+\frac{2p'_\parallel\eta_\nu({p'_\parallel}^2+{p'_\perp}^2)}{M_{Pl}^2})/E'_p+\frac{p'_\parallel+q_\parallel-p}{E'_q}\right|\\
&\cong \left|\frac{3\eta_\nu {p'_\parallel}^2 }{2M_{Pl}^2}+\frac{(\vec{q}_\perp+\vec{p'}_\perp)^2}{2(p-p'_\parallel-q_\parallel)^2}-\frac{{p'_\perp}^2}{2{p'_\parallel}^2}\right|\\
&\cong \frac{3\eta_\nu E^2}{M_{Pl}^2}\left|\left((\frac{1}{2}-\frac{\lambda_{p'}^2}{6x^4})x^2+\frac{\lambda_{q'}^2}{6(1-x-y)^2}\right)\right| \\
&\cong \delta\left|\left((\frac{1}{2}-\frac{\lambda_{p'}^2}{6x^4})x^2+\frac{\lambda_{q'}^2}{6(1-x-y)^2}\right)\right|,
 \end{split}
\label{eqfnl}
\end{equation} 
where we follow Ref.~\cite{matt} and use the longitudinal momentum fractions
$x,\; y$ for $p'$ and $q$ respectively and where we parametrize the
transverse momenta as $\vec{r}_\perp\equiv \hat{r}_\perp\sqrt{\eta_\nu}\frac{p^2}{M_{Pl}}\lambda_r$ for the unit vector $\hat{r}_\perp$
in their direction. By going to the ``preferred frame'' of the incoming $\nu$ we see that the parameters $2\lambda_a$, defined to be non-negative here, are bounded by constants close to $1$.
We have used the 
usual dispersion relations $E_q=\sqrt{q^2+m_e^2},\;E'_q=\sqrt{{q'}^2+m_e^2} $
for the $e^+$ and $e^-$ respectively, where $m_e$ is the rest mass of the electron and have neglected all electron and neutrino masses\footnote{Note that if 
one considers the case in which the $e^-, e^+$ also have the superluminal quadratic dispersive property, with the corresponding parameter $\eta_e$, then our analysis in eq.(\ref{eqfnl}) still holds, with the shift of the RHS to the factor $\delta\left|\left((\frac{1}{2}-\frac{\lambda_{p'}^2}{6x^4})x^2+\frac{\lambda_{q'}^2}{6(1-x-y)^2}-(\eta_e/\eta_\nu)\frac{(1-x-y)^2}{2}\right)\right|$, so that in this case also the RHS of (\ref{mattg2}) should be divided by a factor of $\delta$.}.
We stress that this factor of $\delta$ in (\ref{eqfnl}) should divide the RHS of (\ref{mattg1}) also when one uses it for the process $\nu\rightarrow \nu\nu\bar\nu$ by the standard methods.
\par
We conclude that the estimates of Cohen and Glashow that the 
very energetic $\nu$'s
should lose a large amount of their energy due to Cherenkov $e^+e^-$
radiation hold true for the
observations of OPERA in Ref.~\cite{opera} for superluminal $\nu$'s
with quadratically dispersed values of $\delta$ as well. While showing
this result, we have clarified the issue of the dependence on $\delta$
of this and the attendant related radiative processes.\par
{\bf Note Added in proof.} -- We understand~\cite{matt3} that the difference noted in the text between our Eq.(\ref{eqnew2}) and Eq.(21) in Ref.~\cite{matt2} did not enter the respective results given in Ref.~\cite{matt2}.

\end{document}